\documentclass[12pt,draft]{article}

\usepackage{amsthm,amssymb,amsmath,amscd,amstext,mathrsfs}
\usepackage{latexsym,mathptmx,xypic}

\newtheorem{definition}{Definition}[section]
\newtheorem{theorem}{Theorem}[section]
\newtheorem{lemma}{Lemma}[section]
\newtheorem*{remark}{{\it Remark}}

\DeclareMathAlphabet{\mathpzc}{OT1}{pzc}{m}{it}

\newcommand{\nc}{\newcommand} 

\nc{\C}{{\mathbb C}}
\nc{\R}{{\mathbb R}}
\nc{\HH}{{\mathbb H}}
\nc{\Z}{{\mathbb Z}}
\nc{\N}{{\mathbb N}}
\nc{\dd}{{\rm d}}

\nc{\ii}{{\bf i}}
\nc{\cg}{{\mathscr G}}
\nc{\ch}{{\mathscr H}}

\begin{document}

\title{Exotica or the failure of the strong cosmic censorship\\ 
in four dimensions}

\author{G\'abor Etesi\\
\small{{\it Department of Geometry, Mathematical Institute, Faculty of 
Science,}}\\
\small{{\it Budapest University of Technology and Economics,}}\\
\small{{\it Egry J. u. 1, H \'ep., H-1111 Budapest, Hungary}}
\footnote{e-mail: {\tt etesi@math.bme.hu}}}

\maketitle
                          
\pagestyle{myheadings}
\markright{G. Etesi: Exotica and the strong cosmic censor conjecture}

\thispagestyle{empty}

\begin{abstract} In this letter a generic counterexample to the strong 
cosmic censor conjecture is exhibited. More precisely---taking into 
account that the conjecture lacks any precise formulation yet---first we 
make sense of what one would mean by a ``generic counterexample'' by 
introducing the mathematically unambigous and logically stronger concept 
of a ``robust counterexample''. Then making use of Penrose' nonlinear 
graviton construction (i.e., twistor theory) and a Wick rotation trick we 
construct a smooth Ricci-flat but not flat Lorentzian metric on the 
largest member of the Gompf--Taubes uncountable radial family of large 
exotic $\R^4$'s. We observe that this solution of the Lorentzian vacuum 
Einstein's equations with vanishing cosmological constant provides us with 
a sort of counterexample which is weaker than a ``robust counterexample'' 
but still reasonable to consider as a ``generic counterexample''. It is 
interesting that this kind of counterexample exists only in four 
dimensions. 
\end{abstract}

\centerline{AMS Classification: Primary: 83C75, Secondary: 57N13, 53C28}
\centerline{Keywords: {\it Strong cosmic censor conjecture; Exotic $\R^4$; 
Twistors}}


\section{Introduction}
\label{one}


Certainly one of the deepest open problems of contemporary classical 
general relativity is the validity or invalidity of the {\it strong 
cosmic censor conjecture} \cite{sim-pen}. This is not only a 
technical conjecture of a particular branch of current theoretical 
physics: it deals with the very foundations of our rational description 
of Nature. Indeed, Penrose' original aim in the 1960-70's with formulating this 
conjecture was to protect causality in generic gravitational situations. 
We have the strong conviction that in the {\it classical} physical world at 
least, every physical event (possibly except the 
initial Big Bang) has a physical cause which is another and preceding 
physical event. Since mathematically speaking space-times having this 
property are called {\it globally hyperbolic} our requirement can be 
formulated roughly as follows (cf. e.g. \cite[p. 304]{wal}): 
\vspace{0.1in}

\noindent {\bf SCCC.} {\it A generic (i.e., stable), physically relevant
(i.e., obeying some energy condition) space-time is globally hyperbolic.}
\vspace{0.1in}

\noindent We do not make an attempt here to survey the vast physical and 
mathematical literature triggered by the {\bf SCCC} instead we refer to 
surveys \cite{ise, rin, pen3}. Rather we may summarize the current situation 
as follows. During the course of time the originally single {\bf SCCC} has 
fallen apart into several mathematical or physical 
versions, variants, formulations. For example there exists a generally 
working, mathematically meaningful but from a physical viewpoint rather 
weak version formulated in \cite[p. 305]{wal} and proved in \cite{ete}. 
In another approach to the {\bf SCCC} based on initial value formulation 
\cite[Chapter 10]{wal}, on the one hand, there are certain specific classes of 
space-times in which the {\bf SCCC} allows a mathematically rigorous as well 
as physically contentful formulation whose validity can be established 
\cite{rin}; on the other hand counterexamples to the {\bf SCCC} in this 
formulation also regularily appear in the literature however they are 
apparently too special, not ``generic''. In spite of these sporadic 
counterexamples the overall confidence in the physicist community is 
that an appropriate form of the {\bf SCCC} must hold true hence causality is 
saved.

However we claim to exhibit a {\it generic counterexample to the 
{\bf SCCC}}. What is then the resolution of the apparent contradiction between 
the well-known affirmative solutions and our negative result here? No compact 
smoothable topological $4$-manifold is known carrying only one smooth 
structure. In fact in every well-understood case they admit not only 
more than one but countably infinitely many different smooth structures 
\cite{gom-sti}. In the case of non-compact (relevant for physics) 
topological $4$-manifolds there is even no obstruction against smooth 
structure and they typically accommodate an uncountably family of them 
\cite{gom3}. The astonishing discovery of exotic (or fake) $\R^4$'s 
(i.e., smooth $4$-manifolds which are homeomorphic but not diffeomorphic to 
the usual $\R^4$) by Donaldson, Freedman, Taubes and others in the 1980's 
is just the first example of the general situation 
completely absent in other dimensions. The cases for which the validity of 
the ${\bf SCCC}$ has been verified so far \cite{ise, rin} seem therefore to 
be atypical hence essentially negligable ones; on the contrary, our 
counterexample rests on a typical fake $\R^4$.

The only way to refute the general position adopted here when 
dealing with the ${\bf SCCC}$ was if one could somehow argue that general 
smooth $4$-manifolds are too ``exotic'', ``fake'' or ``weird'' from the 
aspect of physical general relativity. However from the physical viewpoint 
if the ``summing over everything'' approach to quantum gravity is correct 
then very general unconventional but still physical space-times should be 
considered, too \cite{dus}; from the mathematical perspective non-linear 
partial differential equations like Einstein's equations are typically 
also solvable. Consequently it seems that both physically and 
mathematically speaking the true properties of general relativity cannot 
be revealed by understanding it only on simple atypical manifolds; 
the division of smooth $4$-manifolds into ``usual'' and ``unusual'' ones 
can be justified only by conventionalism i.e., one has to evoke historical 
(and technical) arguments to pick up ``usual'' spaces from the bottomless sea 
of smooth $4$-manifolds and abandon others. But looking at things 
optimistically, if it is true that the nature of (quantum) general 
relativity is genuinely not deterministic (as our result suggests) then this 
may open up the exciting possibility that the indeterministic character of 
quantum physics has a quantum gravitational origin. 

Our notational convention throughout the text is that $\R^4$ will denote 
the four dimensional real vector space equipped with its standard 
differentiable manifold structure whilst $R^4$ or $R^4_t$ will denote 
various exotic (or fake) variants. The notation ``$\:\cong\:$'' will 
always mean ``diffeomorphic to'' whilst homeomorphism always will be 
spelled out as ``homeomorphic to''. Finally we note that all set 
theoretical or topological operations (i.e., $\subseteqq$, $\cap$, $\cup$, 
taking open or closed subsets, closures, complements, etc.) will be taken 
in a manifold $M$ with its well-defined standard manifold topology throughout 
the text. In particular for $\R^4$ or the $R^4_t$'s this topology is the 
unique underlying manifold topology.

\vspace{0.1in}

\noindent{\bf Acknowledgement.} Thanks go to B. Kalm\'ar, I. Smoli\'c, 
A. Stipsicz, E. Szab\'o, Sz. Szab\'o and R. Torres for discussions. The work 
was supported in 2014 by the {\it Lend\"ulet} program of the Hungarian 
Academy of Sciences through the ADT Lend\"ulet group at the Alfr\'ed R\'enyi 
Institute of Mathematics.


\section{Definition and construction of a counterexample}
\label{two}


In agreement with the common belief in the physicist and mathematician 
community, formulating the strong cosmic censor conjecture in a 
mathematically rigorous way is obstacled by lacking an overall 
satisfactory concept of ``genericity''. Consequently the 
main difficulty to find a ``generic counterexample'' to the {\bf SCCC} lies 
not in its actual finding (indeed, most of the well-known basic 
solutions of Einstein's equations provide violations of it) but rather in 
proving that the particular counterexample is ``generic''. In this section 
we outnavigate this problem by mathematically formulating the concept 
of a certain counterexample which is logically stronger than a ``generic 
counterexample'' to the {\bf SCCC}. Then we search for a counterexample of 
this kind making use of uncountably many large exotic $\R^4$'s. 

A standard reference here is \cite[Chapters 8,10]{wal}. By a {\it space-time} 
we mean a connected, four dimensional, smooth, time-oriented 
Lorentzian manifold without boundary. By a {\it (continuous) Lorentzian 
manifold} we mean the same thing except that the metric is allowed to be a 
continuous tensor field only.

\begin{definition} Let $(S, h, k)$ be an initial data set for Einstein's 
equations with $(S, h)$ a complete Riemannian $3$-manifold and with a 
fundamental matter represented by a stress-energy tensor $T$ obeying the 
dominant energy condition. Let $(D(S), g\vert_{D(S)})$ be the unique maximal 
Cauchy development of this initial data set. Let $(M,g)$ be a further maximal 
extension of $(D(S), g\vert_{D(S)})$ as a (continuous) Lorentzian manifold if 
exists. That is, $(D(S), g\vert_{D(S)})\subseteqq (M,g)$ is a (continuous) 
isometric embedding which is proper if $(D(S), g\vert_{D(S)})$ is 
still extendible and $(M,g)$ does not admit any further proper isometric 
embedding. (If the maximal Cauchy development is inextendible then 
put simply $(M,g):=(D(S), g\vert_{D(S)})$ for definiteness.)

The (continuous) Lorentzian manifold $(M',g')$ is a {\rm perturbation of 
$(M,g)$ relative to $(S,h,k)$} if 
\begin{itemize}

\item[{\rm (i)}] $M'$ has the structure 
\[M':=\mbox{{\rm the connected component of $M\setminus\ch$ containing $S$}}\]
where, for a connected open subset $S\subset U\subseteqq M$ 
containing the initial surface, $\emptyset\subseteqq\ch\subseteqq\partial U$ 
is a closed subset in the boundary of $U$ (consequently $M'$ is open in $M$ 
hence inherits a differentiable manifold structure);

\item[{\rm (ii)}] $g'$ is a solution of Einstein's 
equations at least in a neighbourhood of $S\subset M'$ with a fundamental 
matter represented by a stress-energy tensor $T'$ obeying the dominant energy 
condition at least in a neighbourhood of $S\subset M'$; 

\item[{\rm (iii)}] $(M',g')$ does not admit further proper isometric embeddings 
and $(S,h')\subset (M',g')$ with $h':=g'\vert_S$ is a spacelike complete 
sub-$3$-manifold. 
\end{itemize}

\label{perturbacio}
\end{definition}

\begin{remark}\rm 1. It is crucial that in the spirit of relativity theory 
we consider metric perturbations of the {\it four} dimensional space-time 
(whilst keeping its underlying smooth structure fixed)---and not those of a 
{\it three} dimensional initial data set. This natural class of perturbations 
is therefore immense: it contains all connected manifolds containing the 
initial surface but perhaps being topologically different from the original 
manifold. The perturbed metric is a physically relevant 
solution of Einstein's equations at least in the vicinity of $S\subset M'$ 
such that $(M',g')$ is inextendible and $(S,h')\subset (M',g')$ is still 
spacelike and complete. In other words these perturbations are physical 
solutions allowed to blow up along closed `` boundary subsets'' 
$\emptyset\subseteqq\ch\subset M$; the notation $\ch$ for these subsets 
indicates that among them the (closure of the) Cauchy horizon $H(S)$ of 
$(S,h,k)$ may also appear. Beyond the non-singular perturbations with 
$\ch=\emptyset$ of any space-time a prototypical example with 
$\ch\not=\emptyset$ is the physical perturbation $(M',g')$ of the 
(maximally extended) undercharged Reissner--Nordstr\"om space-time $(M,g)$ 
by taking into account the full backreaction of a pointlike particle or any 
classical field put onto the originally pure electro-vacuum space-time 
(``mass inflation''). In this case the singularity subset $\ch$ is expected to 
coincide with the (closure of the) full inner event horizon of the 
Reissner--Nordstr\"om black hole which is the Cauchy horizon for the standard 
initial data set inside the maximally extended space-time \cite{sim-pen}. 
A similar perturbation of the Kerr--Newman space-time is another example 
with $\ch\not=\emptyset$.

2. Accordingly, notice that in the above definition of perturbation none 
of the terms ``generic'' or ``small'' have been used. This indicates 
that if such types of perturbations can be somehow specified then one 
should be able to recognize them among the very general but still physical 
perturbations of a space-time as formulated in Definition \ref{perturbacio}. 
\end{remark}

\noindent Now we are in a position to formulate in a mathematically 
precise way what we mean by a ``robust counterexample'' to the {\bf SCCC} 
as formulated roughly in the Introduction. 

\begin{definition} Let $(S, h, k)$ 
be an initial data set for Einstein's equations with $(S, h)$ a complete 
Riemannian $3$-manifold and with a fundamental matter represented by a 
stress-energy tensor $T$ obeying the dominant energy condition. 
Assume that the maximal Cauchy development of this initial data set is 
extendible i.e., admits a (continuous) isometric embedding as a proper open 
submanifold into an inextendible (continuous) Lorentzian manifold $(M,g)$.

Then $(M,g)$ is a {\rm robust counterexample to the {\bf SCCC}} 
if it is very stably non-globally hyperbolic i.e., all of its perturbations 
$(M',g')$ relative to $(S,h,k)$ are not globally hyperbolic.

\label{ellenpelda}
\end{definition}

\begin{remark}\rm 1. Concerning its logical status it is reasonable to 
consider this as a {\it generic} counterexample because the perturbation 
class of Definition \ref{perturbacio} is expected to contain all ``generic 
perturbations'' whatever they are. Consequently in Definition \ref{ellenpelda} 
we are dealing with a stronger statement than the logical negation of the 
affirmative sentence in {\bf SCCC}. 

2. The trivial perturbation i.e., the extension $(M,g)$ itself in 
Definition \ref{ellenpelda} cannot be globally hyperbolic as observed already 
in \cite[Remark after Theorem 2.1]{ete}. 
\end{remark}

\noindent Strongly influenced by \cite{ber-san, che-nem} we take now an 
excursion into the weird world of four dimensional exotic {\it m\'enagerie} 
(or rather {\it plethora}) in order to attack the {\bf SCCC}. 
A standard reference here is \cite[Chapter 9]{gom-sti}. Our construction 
is based on a specific fake $\R^4$ whose proof of existence is very 
involved: it is based on the works of Gompf \cite{gom1, gom2} and Taubes 
\cite{tau}. Those properties of this exotic $\R^4$ which will be used here 
are summarized as follows (cf. \cite[Lemma 9.4.2, Addendum 9.4.4 and 
Theorem 9.4.10]{gom-sti}):

\begin{theorem} There exists a pair $(R^4,K)$ consisting of a 
differentiable $4$-manifold $R^4$ homeomorphic but not diffeomorphic to the 
standard $\R^4$ and a compact oriented smooth $4$-manifold $K\subset R^4$ 
such that 
\begin{itemize}

\item[{\rm (i)}] $R^4$ cannot be smoothly embedded into the standard 
$\R^4$ i.e., $R^4\not\subseteqq\R^4$ but it can be smoothly embedded as a 
proper open subset into the complex projective plane i.e., 
$R^4\subsetneqq\C P^2$;

\item[{\rm (ii)}] Take a homeomorphism $f:\R^4\rightarrow R^4$, let 
$0\in B^4_t\subset\R^4$ be the standard open $4$-ball of radius $t\in\R^+$ 
centered at the origin and put $R^4_t:=f(B^4_t)$ and 
$R^4_{+\infty}:=R^4$. Then 
\[\left\{ R^4_t\:\left\vert\:\mbox{$r\leqq t\leqq +\infty$ such that 
$0<r<+\infty$ satisfies $K\subset R^4_r$}\right.\right\}\]
is an uncountable family of nondiffeomorphic exotic $\R^4$'s none of them 
admitting a smooth embedding into $\R^4$ i.e., $R^4_t\not\subseteqq\R^4$ 
for all $r\leqq t\leqq +\infty$. 

\end{itemize}

\noindent In what follows this family will be referred to as the {\rm 
radial family of large exotic $\R^4$'s}. $\Diamond$

\label{egzotikusnagycsalad}
\end{theorem}

\begin{remark}\rm From Theorem \ref{egzotikusnagycsalad} we deduce that 
for all $r<t<+\infty$ there is a sequence of smooth proper embeddings 
\[R^4_r\subsetneqq R^4_t\subsetneqq R^4_{+\infty}=R^4\subsetneqq\C P^2\] 
which are very wild: the complement $\C P^2\setminus R^4$ is homeomorphic 
to $S^2$ (regarded as an only continuously embedded projective line in the 
projective plane) consequently it does not contain any open $4$-ball in 
$\C P^2$; hence in particular if $\C P^2=\C^2\cup\C P^1$ is any holomorphic 
decomposition then $R^4\cap\C P^1\not=\emptyset$ (because otherwise 
$R^4\subseteqq\C^2\cong\R^4$ would hold, a contradiction). This 
demonstrates that the members of the large radial family live ``somewhere 
between'' $\C^2$ and its projective closure $\C P^2$. However a more 
precise identification or location of them is a difficult task because 
these large exotic $\R^4$'s---although being honest differentiable 
$4$-manifolds---are very transcendental objects \cite[p. 366]{gom-sti}: 
they require infinitely 
many $3$-handles in any handle decomoposition (like any other known large 
exotic $\R^4$) and there is presently\footnote{More precisely in the year 
1999, cf. \cite{gom-sti}.} no clue as how one might draw explicit handle 
diagrams of them (even after removing their $3$-handles). We note that the 
structure of small exotic $\R^4$'s i.e., which admit smooth embeddings 
into $\R^4$, is better understood, cf. \cite[Chapter 9]{gom-sti}. 
\end{remark}

\noindent We proceed further and construct a solution of the Lorentzian 
vacuum Einstein's equations (with vanishing cosmological constant) on 
$R^4$ by the aid of Penrose' nonlinear graviton construction 
\cite{pen2}.

\begin{theorem} The space $R^4$ from Theorem \ref{egzotikusnagycsalad} 
carries a smooth Lorentzian Ricci-flat metric $g$. Moreover there exists 
an open (i.e., non-compact without boundary) contractible spacelike and 
complete sub-$3$-manifold $(S,h)\subset(R^4, g)$ in it such that $h=g\vert_S$. 

The Ricci-flat Lorentzian manifold $(R^4, g)$ might be timelike and (or) 
null geodesically incomplete.

\label{tvisztor}
\end{theorem}

\noindent {\it Proof.} The proof consists of two steps: (i) we construct 
a {\it Riemannian} Ricci-flat metric on $R^4$ via twistor theory by 
exploiting the embedding $R^4\subset\C P^2$; (ii) ``Wick rotate'' this 
solution into a {\it Lorentzian} one by exploiting the 
contractibility of $R^4$.

(i) The original nonlinear graviton construction of Penrose 
\cite{pen2}, as summarized very clearly in \cite{hit1} or 
\cite[\S 4]{hit2}, consists of the following data: 
\begin{itemize}

\item[$*$] A complex $3$-manifold $Z$, the total space of 
a holomorphic fibration $\pi :Z\rightarrow\C P^1$;

\item[$*$] A complex $4$-paremeter family of sections $Y\subset Z$, each 
with normal bundle $H\oplus H$ (here $H$ is the dual of the tautological 
bundle i.e., the unique holomorphic line 
bundle on $Y\cong\C P^1$ with $\langle c_1(H), [Y]\rangle =1$);

\item[$*$] A non-vanishing holomorphic section $s$ of 
$K_Z\otimes\pi^*H^4$ (here $K_Z$ is the canonical bundle of $Z$);

\item[$*$] A real structure $\tau :Z\rightarrow Z$ such that $\pi$ and $s$ 
are compatible and $Z$ is fibered by the real sections of the family 
(here $\C P^1$ is given the real structure of the antipodal map 
$u\mapsto -\overline{u}^{-1}$). 

\end{itemize}

\noindent These data allow one to construct a Ricci-flat and self-dual 
(i.e., the Ricci tensor and the antiself-dual part of the Weyl tensor 
vanishes) solution $(M,g)$ of the {\it Riemannian} Einstein's vacuum 
equations (with vanishing cosmological constant) in a well-known way. 
The holomorphic lines $Y\subset Z$ form a locally complete family and 
fit together into a complex $4$-manifold $M^\C$. This space carries a 
natural complex conformal structure by declaring two nearby points $y',y''\in 
M^\C$ to be null-separated if the corresponding lines intersect i.e., 
$Y'\cap Y''\not=\emptyset$ in $Z$. Infinitesimally this means that on 
every tangent space $T_yM^\C =\C^4$ a null cone is specified. 
Restricting this to the real lines parameterized by an embedded real 
$4$-manifold $M\subset M^\C$ we obtain the real conformal class $[g]$ of a 
Riemannian metric on $M$. The isomorphism $s: K_Z\cong\pi^*H^{-4}$ is 
essentially uniquely fixed by its holomorphic and reality properties and 
gives rise to a volume form on $M$ this way fixing the metric $g$ in the 
conformal class. Since $Z$ can be identified with the projective negative 
chiral spinor bundle $P\Sigma^-$ over $M$ we obtain a smooth twistor 
fibration $p:Z\rightarrow M$ whose fibers are $\C P^1$'s hence 
$\pi :Z\rightarrow\C P^1$ can be regarded as a parallel translation with 
respect to a flat connection which is nothing but the induced negative spin 
connection of $g$ on $\Sigma^-$. This partial flatness implies that $g$ is 
Ricci-flat and self-dual. For more details cf. \cite{hit1, hit2}.

In the case of our large exotic $\R^4$ i.e., $R^4$ from Theorem 
\ref{egzotikusnagycsalad} these data arise as follows. Putting 
$Z:=P(T^*\C P^2)$ to be the projective cotangent bundle we obtain the 
twistor fibration $p:Z\rightarrow\C P^2$ of the complex projective 
space. More classically $Z$ can be viewed as the flag manifold 
$F_{12}(\C ^3)$ consisting of pairs $({\mathfrak l},{\mathfrak p})$ 
where $0\in{\mathfrak l}\subset\C^3$ is a line and ${\mathfrak l} 
\subset{\mathfrak p}\subset\C^3$ is a plane containing the line. The 
projection sends $({\mathfrak l},{\mathfrak p}) \in F_{12}(\C ^3)\cong 
Z$ into $[\:{\mathfrak l}\:]\in\C P^2$. This is a smooth $\C 
P^1$-fibration over $\C P^2$. Part (i) of Theorem 
\ref{egzotikusnagycsalad} tells us that $R^4\subset\C P^2$. Writing 
$Z':=Z\vert_{R^4}$ and $p':=p\vert_{Z'}$ the restricted twistor 
fibration $p':Z'\rightarrow R^4$ is topologically trivial i.e., $Z'$ is 
homeomorphic to $R^4\times S^2\cong\R^4\times S^2$ because $R^4$ is 
contractible.\footnote{The full twistor fibration $p:Z\rightarrow\C P^2$ 
is non-trivial.} Take an $({\mathfrak l},{\mathfrak p})\in Z'$ and 
choose a local complex coordinate $z\in\C$ on the corresponding 
projective line $[{\mathfrak p}]\subset\C P^2$ such that the point 
$[\:{\mathfrak l}\:]\in [{\mathfrak p}]$ satisfies $z([\:{\mathfrak l}\:])=0$. 
Define the unique element $[{\mathfrak l}_\infty]\in [{\mathfrak p}]$ by the 
two infima 
\[\left\vert z([\:{\mathfrak l''}\:])\right\vert := 
\inf\limits_{[{\mathfrak l}']\in [{\mathfrak p}]\cap (\C P^2\setminus 
R^4)} \vert z([\:{\mathfrak l}'\:])\vert\in [0,+\infty )\:\:\:, 
\:\:\:\:\:\:\:\:\:\: {\rm arg}\:z([{\mathfrak l}_\infty ]):= 
\inf\limits_{[{\mathfrak l}'']\in [{\mathfrak p}]\cap (\C P^2\setminus 
R^4)}\: {\rm arg}\:z([\:{\mathfrak l}''\:])\in [0,2\pi )\:\:\:.\] The 
assignment $[{\mathfrak p}]\mapsto [{\mathfrak l}_\infty ]\in\C P^2 
\setminus R^4$ always satisfies $[{\mathfrak l}_\infty ]\not= 
[\:{\mathfrak l}\:]$ because $[\:{\mathfrak l}\:]\in R^4$ is an inner point. 
Pick any projective line 
$\ell\subset\C P^2\setminus\{[\:{\mathfrak l}\:]\}$ and identify 
$(\C P^2\setminus\{[\:{\mathfrak l}\:]\}\:,\:\ell)$ with the line bundle 
$(H, \C P^1)$ such that $\C P^2\setminus\{[\:{\mathfrak l}\:]\}$ is the total 
space $H$, $\ell$ is the zero section hence the base $\C P^1$ and the 
punctured projective lines $[{\mathfrak p}]\setminus\{[\:{\mathfrak l}\:]\}
\subset\C P^2\setminus\{[\:{\mathfrak l}\:]\}$ through 
$[\:{\mathfrak l}\:]\in\C P^2$ represent the fibers of the line bundle. This 
way we can regard the assignment 
$[{\mathfrak p}]\mapsto [{\mathfrak l}_\infty ]$ 
as a section $s:\ell\rightarrow \C P^2\setminus\{[\:{\mathfrak l}\:]\}$. 
The Fubini--Study metric gives rise to a fiberwise Hermitian metric on 
$\C P^2\setminus\{[\:{\mathfrak l}\:]\}$ turning $s$ into an $L^2$-section. 
Its $L^2$-orthogonal projection onto the $2$ dimensional subspace of 
holomorphic sections gives rise to a unique holomorphic section what we 
continue to write as 
$s:\ell\rightarrow \C P^2\setminus\{[\:{\mathfrak l}\:]\}$. Consequently this 
holomorphic section yields a well-defined {\it holomorphic assignment} what 
we denote by $[{\mathfrak p}]\mapsto [{\mathfrak l}_\infty ]$ 
as before.\footnote{However observe that after the $L^2$-projection both 
$[{\mathfrak l}_\infty ]\in\C P^2\setminus R^4$ or 
$[{\mathfrak l}_\infty ]\in R^4$ can occur.} 
Fix a point $[{\mathfrak l}_0]\in R^4$ and using this assignment define 
$\pi ':Z'\rightarrow p'^{-1}([{\mathfrak l}_0])\cong\C P^1$ by putting 
\[\pi '(({\mathfrak l},\!{\mathfrak p})):= 
\mbox{$({\mathfrak l}_0, \!{\mathfrak p}_0)$ where 
${\mathfrak p}_0 \supset {\mathfrak l}_0$ in $\C^3$ satisfies that 
$[{\mathfrak p}_0]\subset\C P^2$ connects $[{\mathfrak l}_0]$ with 
$[{\mathfrak l}_\infty ]$ assigned to $[{\mathfrak p}]$}\] 
(see Fig. $1$.). By construction this is a holomorphic map whose fibers are 
diffeomorphic to $R^4$. Then restricting everything onto $R^4\subset \C P^2$ it
is immediate that $Z'$ contains the complex $4$-parameter family $(R^4)^\C$ of
holomorphic lines and the corresponding real lines parameterize $R^4$. The map
$\pi '$ is compatible with the real structure. A non-vanishing holomorphic real
section $s$ of $K_{Z'}\otimes\pi '^*H^4$ then fixes the {\it Riemannian} metric
$g_1$ on $R^4$ which is Ricci-flat.
\vspace{0.1in}

\centerline{
\setlength{\unitlength}{1cm}
\begin{picture}(5,3)
\thicklines
\put(1,2.5){\circle*{0.2}}
\put(1,1){\circle*{0.2}}
\put(4,1){\circle*{0.2}}
\put(1,0){\line(0,1){3}}
\put(0,1){\line(1,0){5}}
\put(0.3,2.5){$[\:{\mathfrak l}\:]$}
\put(0.3,0.5){$[{\mathfrak l}_\infty ]$}
\put(4,0.5){$[{\mathfrak l}_0]$}
\put(1.2,1.7){$[{\mathfrak p}]$}
\put(2.5,1.3){$[{\mathfrak p}_0]$}
\end{picture}
            }
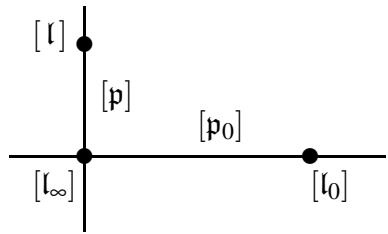
\centerline{Figure 1. Construction of the map $\pi '$.}
\vspace{0.1in}

We proceed further and demonstrate that $(R^4,g_1)$ is complete. 
Since both the Fubini--Study metric $g_0$ and this Ricci-flat metric $g_1$ 
stem from the same complex structure on the twistor space we know from 
twistor theory that these metrics are in fact conformally equivalent. 
Therefore there exists a smooth non-constant positive function 
$\varphi :R^4\rightarrow\R^+$ such that 
$\varphi^{-2}\cdot\left( g_0\vert_{R^4}\right) = g_1$ and satisfying with 
respect to $g_0$ the following equations:
\begin{equation}
\left\{\begin{array}{ll}
\Delta\varphi^{-1}+\frac{1}{6}{\rm scal}(g_0\vert_{R^4})\varphi^{-1}&=0
\:\:\:
\mbox{(vanishing of the scalar curvature of $g_1$)};\\
\nabla^2\varphi +\frac{1}{4}\Delta\varphi\cdot (g_0\vert_{R^4})&= 0\:\:\:
\mbox{(vanishing of the traceless Ricci tensor of $g_1$)}.
\end{array}\right.
\label{skalazas}
\end{equation}
Taking into account that the scalar curvature of the Fubini--Study metric is 
constant it follows from the first equation of (\ref{skalazas}) and the 
maximum principle that $\varphi^{-1}$ diverges along $\C P^2\setminus R^4$ 
hence on the one hand the function $\log \varphi^{-1}:R^4\rightarrow\R$ 
is a proper function on $R^4$. 
 
Denoting by $X$ the dual vector field to $\dd\varphi$ with respect to 
$g_0$ the decomposition of the $(0,2)$-type symmetric 
tensor field $\nabla^2\varphi$ into trace and traceless symmetric parts gives  
\[\nabla^2\varphi +\frac{1}{4}\Delta\varphi\cdot (g_0\vert_{R^4})
=\frac{1}{2}\left( L_X(g_0\vert_{R^4})+\frac{1}{2}\Delta\varphi\cdot  
(g_0\vert_{R^4})\right)\]
hence the second equation of (\ref{skalazas}) says that $X$ (or $\varphi$) 
satisfies the {\it conformal Killing equation} with respect to the (restricted)
Fubini--Study metric: $L_X(g_0\vert_{R^4})+\frac{1}{2}\Delta\varphi\cdot 
(g_0\vert_{R^4})=0$. 
The conformal Killing equation for $X$ can be prolongated in a well-known way 
i.e., can be re-written in terms of the conformal Killing data 
$(\dd\varphi\:,\:\dd^2\varphi = 0\:,\:-\Delta\varphi\:,\:
-\dd (\Delta\varphi ) )$ for $X$ as a system of differential equations 
(cf. \cite[Eqn. B.3]{ger}). The relevant equation for us deals with the 
fourth data and on the Einstein manifold $(R^4, g_0\vert_{R^4})$ with 
constant scalar curvature takes the shape $\nabla (\dd (\Delta\varphi ))=
-\frac{1}{12}{\rm scal}(g_0\vert_{R^4})\Delta\varphi\cdot (g_0\vert_{R^4})$. 
Combining this with the second equation of (\ref{skalazas}) leads to
\[\nabla\left(\dd\left(\Delta\varphi -\frac{1}{3}{\rm scal}(g_0\vert_{R^4})
\varphi\right)\right) =0\:\:\:.\] 
The restricted Fubini--Study geometry is still irreducible in the sense that 
its holonomy group acts irreducibly on the tangent spaces hence we conclude 
that in fact 
$\dd\left(\Delta\varphi -\frac{1}{3}{\rm scal}(g_0\vert_{R^4})
\varphi\right)=0$ i.e., there exists $c_1\in\R$ such that 
\begin{equation}
\Delta\varphi =\frac{1}{3}{\rm scal}(g_0\vert_{R^4})\varphi +c_1
\label{laplace}
\end{equation} 
holds. It again follows from the maximum principle that surely $c_1\not= 0$. 
Adjusting the standard identity $0=(\Delta\varphi )\varphi^{-1} +
2g_0(\dd\varphi\:,\dd\varphi^{-1})+ \varphi\Delta\varphi^{-1}$ as 
$\varphi^2\vert\dd\varphi^{-1}\vert^2_{g_0}=\frac{1}{2}(
\varphi\Delta\varphi^{-1}+\varphi^{-1}\Delta\varphi )$, plugging the 
first equation of (\ref{skalazas}) as well as (\ref{laplace}) into it and 
carefully writing $\vert\xi\vert_{g_1}=\varphi\vert\xi\vert_{g_0}$ on 
$1$-forms we obtain on the other hand the estimate  
\[\vert\dd (\log \varphi^{-1})\vert^2_{g_1}
=\varphi^4\vert\dd\varphi^{-1}\vert^2_{g_0}=
\frac{1}{2}\left(\varphi^3\Delta\varphi^{-1} 
+\varphi\Delta\varphi\right) = 
\frac{1}{12}{\rm scal}(g_0\vert_{R^4})\varphi^2+
\frac{1}{2}c_1\varphi\leqq c_2\]
with some $c_2\in\R^+$ because ${\rm scal}(g_0\vert_{R^4})$ is constant and 
$\varphi$ is bounded. Recalling a classical result of Gordon \cite{gor} a 
Riemannian manifold is complete if and only if it admits an at least $C^3$ 
proper function whose gradient is bounded in modulus. Since 
$\log\varphi^{-1}:R^4\rightarrow\R$ satisfies these conditions we conclude 
that the Ricci flat space $(R^4,g_1)$ is moreover complete.\footnote{The 
metric $g_1$ is additionally self-dual and $R^4$ is simply connected hence 
$(R^4, g_1)$ is in fact a {\it hyper-K\"ahler gravitational instanton}. 
Therefore this geometry is expected to make a dominant contribution to the 
Euclidean quantum gravitational partition function, cf. \cite{dus}.} 

(ii) Next we ``Wick rotate'' this Riemannian solution into a Lorentzian 
one. We begin with the construction of a Riemannian sub-$3$-manifold 
$(S,h)\subset (R^4,g_1)$. The boundary of the unit disk bundle 
inside the total space of the line bundle $H$ over $\C P^1$ is a circle 
bundle over its zero section $\C P^1$ more precisely a 
Hopf fibration hence is a $3$-manifold homeomorphic to $S^3$. Fixing an 
$[{\mathfrak l}_\infty ]\in\C P^2\setminus R^4$ we identify again the 
total space $H$ with $\C P^2\setminus\{ [{\mathfrak l}_\infty ]\}$ 
and denote by $N\subset\C P^2\setminus\{ [{\mathfrak l}_\infty ]\}$ the 
image of this $3$-manifold. Define 
\[S:=\mbox{one connected component of $N\cap R^4$}\:\:\:.\] 
Every exotic $\R^4$ in general hence our $R^4$ in particular, has the property 
that it contains a compact subset $C\subset R^4$ which cannot be surrounded 
by a smoothly embedded $S^3\subset R^4$ \cite[Exercise 9.4.1]{gom-sti}. 
Taking the radii of the constituent circles of $N$ sufficiently 
large we can suppose by the compactness of $C$ that $C\cap S=\emptyset$ i.e., 
$S$ could surround $C$ if $S$ was homeomorphic to $S^3$. This would be a 
contradiction hence $S\subset R^4$ is an open (i.e., non-compact without 
boundary) and connected sub-$3$-manifold of $R^4$. Therefore, exploiting the 
contractibility of $R^4$ we conclude that $S$ is an open contractible 
sub-$3$-manifold within $R^4$. Putting $h:=g_1\vert_S$ we therefore 
obtain an open contractible Riemannian sub-$3$-manifold 
$(S,h)\subset (R^4,g_1)$ which is complete. 

Consider $(S,h)\subset (R^4,g_1)$ constructed above. Pick a real 
line bundle $L$ over $R^4$ such that 
$L\subset TR^4$ and its orthogonal complement $L^\perp\subset TR^4$ within 
$TR^4$ satisfies $L^\perp\vert_S\cong TS$. Moreover take the complex tangent 
bundle $T(R^4)^\C$ of the complexification $(R^4)^\C$ and restrict it onto 
$R^4\subset (R^4)^\C$. This $T(R^4)^\C\vert_{R^4}$ is a trivial smooth 
rank-$4$ complex vector bundle over $R^4$ and obviously contains the 
real tangent bundle $TR^4$. Consider the imaginary line bundle $\ii L
\subset T(R^4)^\C\vert_{R^4}$. We claim that $L$ can be fiberwisely 
rotated into $\ii L$ within $T(R^4)^\C\vert_{R^4}$ in a continuous 
manner over the whole $R^4$. To see this let $\cg$ denote the gauge 
group of the complex vector bundle $T(R^4)^\C\vert_{R^4}$ consisting of 
smooth fiberwise ${\rm SO}(4,\C )$-transformations (provided by the 
complexification of the metric constructed on $R^4$ by twistor theory). Assume 
that an element $\alpha\in\cg$ satisfies 
\[\alpha L=\ii L\:\:\:.\] 
Then if $\beta_L\in\cg$ and $\beta_{\ii L}\in\cg$ are rotations fixing $L$ 
and $\ii L$ within $T(R^4)^\C\vert_{R^4}$ respectively then 
\[(\beta^{-1}_{\ii L}\alpha \beta_L)L=\ii L\] 
holds as well. The fiberwise stabilizers of 
both $L$ and $\ii L$ are isomorphic to $\Z_2\subset{\rm SO}(4,\C )$ therefore 
the existence of an $\alpha\in\cg$ implies that a principal 
$\Z_2\times\Z_2$-bundle $P_{L,\ii L}$ over $R^4$ (within the 
${\rm SO}(4,\C )$-bundle providing the gauge group $\cg$) given by the 
relative positions of $L$ and $\ii L$ within $T(R^4)^\C\vert_{R^4}$ admits a 
continuous section $(\beta_L, \beta_{\ii L}) \in C^0(R^4; P_{L,\ii L})$. 
Consequently this principal bundle, the ``obstruction bundle'' of the rotation, 
must be trivial otherwise $\alpha\in\cg$ cannot exists. Standard obstruction 
theory says that the only obstruction class against $P_{L,\ii L}$ to be 
trivial lives in the cohomology group 
\[H^1(R^4\:;\:\pi_0(\Z_2\times\Z_2))\cong H^1(R^4\:;\:\Z_2)\times 
H^1(R^4\:;\:\Z_2)\:\:\:.\] 
However referring to the contractibility of $R^4$ once again we conclude that 
$H^1(R^4\:;\:\Z_2)=0$ hence the continuous ``Wick rotation'' of 
$L\subset TR^4$ into $\ii L\subset T(R^4)^\C\vert_{R^4}$ can be performed.

The real structure on $Z'$ cuts out $R^4\subset (R^4)^\C$ and its 
infinitesimal form at $x\in R^4\subset (R^4)^\C$ gives rise to a real 
subspace $\R^4=T_xR^4\subset T_x(R^4)^\C =\C^4$; in addition twistor theory 
equips $T_xR^4$ with a real scalar product $(g_1)_x$, too. Taking the 
complex linear extension of this real scalar product we obtain a complex 
scalar product on $T_x(R^4)^\C$ yielding an inclusion of the corresponding 
spin groups 
\[{\rm Spin}(4)\cong{\rm SU}(2)\times{\rm SU}(2)
\subset{\rm SL}(2,\C )\times{\rm SL}(2,\C)\cong {\rm Spin}(4,\C )\:\:\:.\] 
Since $T_xR^4=L_x\oplus L^\perp_x$ the complex scalar product restricted to 
$\R^4 =\ii L_x\oplus L^\perp_x\subset T_x(R^4)^\C =\C^4$ 
gives an indefinite real scalar product with its associated real spin group 
\[{\rm Spin}(3,1)\cong{\rm SL}(2,\C )\subset {\rm SL}(2,\C )
\times{\rm SL}(2,\C)\cong {\rm Spin}(4,\C )\] 
being diagonally embedded into the complex spin group. Therefore, on the one 
hand, the real rank-$4$ bundle 
$\ii L\oplus L^\perp\subset T(R^4)^\C\vert_{R^4}$ 
over $R^4$ carries a metric $g$ which is Lorentzian and continues to be 
Ricci-flat but not flat because it follows from the above analysis of the 
spin groups that its full Weyl tensor is not zero. On the other hand 
$\ii L\oplus L^\perp$ can be identified with $TR^4$ because both bundles are 
trivial. Consequently we obtain a {\it Lorentzian} Ricci-flat manifold 
what we call $(R^4,g)$. It also possesses a complete spacelike 
$(S,h)\subset (R^4,g)$ which is nothing but the previously constructed 
$(S,h)\subset (R^4, g_1)$.

We conclude that $R^4$ admits a solution of the Lorentzian vacuum 
Einstein's equations as desired. However this solution might be incomplete in 
the non-spacelike directions. $\Diamond$
\vspace{0.1in}

\noindent After this technical warm-up we inspect $(R^4,g)$ concerning its 
global hyperbolicity. 

\begin{lemma} Consider the pair $(R^4, K)$ from Theorem 
\ref{egzotikusnagycsalad} and the Ricci-flat Lorentzian manifold $(R^4,g)$ of 
Theorem \ref{tvisztor} with its open contractible spacelike and complete 
sub-$3$-manifold $(S,h)\subset (R^4,g)$. Let $(S,h,k)$ be the initial data 
set inside $(R^4,g)$ induced by $(S,h)$ and let $(M',g')$ be a perturbation of 
$(R^4,g)$ relative to $(S,h,k)$ as in Definition \ref{perturbacio}. 

Assume that $K\subset M'$ holds. Then $(M',g')$ is not globally hyperbolic. 

\label{ellenpeldalemma}
\end{lemma}

\noindent {\it Proof.} First we prove that the trivial perturbation 
i.e., $(R^4, g)$ itself is not globally hyperbolic. Since $R^4$ is an 
exotic $\R^4$ then by a result of McMillen \cite{mcm} it does not 
admit a smooth splitting like $R^4\cong W\times\R$ where $W$ is an open 
contractible $3$-manifold. Hence it follows from the smooth splitting 
theorem for globally hyperbolic space-times \cite{ber-san} that 
$(R^4,g)$ cannot be globally hyperbolic.\footnote{Or simply we can refer 
to \cite[Theorem A]{che-nem} to get this result.} Consequently the 
initial data set $(S,h,k)$ induced by $(S,h)\subset (R^4,g)$ is only a 
partial initial data set inside $(R^4,g)$.

Let us secondly consider its non-trivial perturbations $(M',g')$ relative to 
$(S,h,k)$ satisfying $K\subset M'$. Suppose that $(M',g')$ is globally 
hyperbolic. Referring to Definition \ref{perturbacio} we know that 
$(S,h')\subset (M',g')$ is a complete spacelike submanifold hence we can use 
it to obtain an initial data set $(S,h',k')$ for $(M',g')$. Again 
by \cite{ber-san} we find $M'\cong S\times\R$ and taking into account that 
$S$ is an open contractible manifold we can refer again to \cite{mcm} to 
conclude that $M'\cong\R^4$. Essentially by Uryshon's lemma we can find in 
part (ii) of Theorem \ref{egzotikusnagycsalad} a homeomorphism 
$f:\R^4\rightarrow R^4$ and a value $r\leqq t_0\leqq +\infty$ such that 
with the corresponding exotic space $R^4_{t_0}=f(B^4_{t_0})$ a sequence of 
smooth embeddings
\[\begin{matrix}
K&\!\!\!\subsetneqq\!\!\!&R^4_{t_0}&\!\!\!\subseteqq&M'&\!\!\!\subseteqq 
\:\:\:R^4_{+\infty}& =& R^4\\
           &            &         &           &\wr\!\!\parallel&  & \\
           &            &         &           &\R^4              &  &
   \end{matrix}\] 
exists. However this is a contradiction because $R^4_{t_0}$ is a member of 
the radial family of large exotic $\R^4$'s of Theorem 
\ref{egzotikusnagycsalad} consequently it cannot be smoothly embedded into 
$M'\cong\R^4$. This demonstrates that our supposition was wrong hence 
$(M',g')$ cannot be globally hyperbolic as well. $\Diamond$
\vspace{0.1in}

\begin{remark}\rm The simple assumption $K\subset M'$ says that the 
perturbation about $S\subset R^4$ is large enough in the topological 
sense hence is capable to ``scan'' the exotic regime of $R^4$. In fact 
this condition is effectively necessary to exclude globally hyperbolic 
perturbations of $(R^4,g)$. Taking $M':=N_\epsilon (S)\subset R^4$ to be 
a small tubular neighbourhood of $S\subset R^4$ then the contractibility 
of $S$ implies $N_\varepsilon (S)\cong S\times\R$ hence again by 
\cite{mcm} we know that $N_\varepsilon (S)\cong\R^4$. Therefore putting 
$g'$ just to be the standard Minkowski metric on $M'$ then $(M',g')$ is the 
usual Minkowski space-time hence is a globally hyperbolic perturbation of 
$(R^4,g)$ relative to $(S,h,k)$. This perturbation is ``small'' in the 
topological sense above however might be ``large'' in any analytical 
sense i.e., the corresponding $(S,h',k')$ might siginificantly deviate from 
the original $(S,h,k)$. 
\end{remark}

\noindent Taking into account that the class of perturbations $(M',g')$ of 
$(R^4,g)$ relative to $(S,h,k)$ has to satisfy a non-trivial condition 
$K\subset M'$ in order to be non-globally hyperbolic the space $(R^4,g)$ 
is not a robust counterexample to the {\bf SCCC} in the strict sense of 
Definition \ref{ellenpelda}. However this condition is just a mild 
topological one hence the corresponding perturbation class is certainly 
still enormously vast. Therefore in our opinion it is reasonable to 
say that {\it the Ricci-flat Lorentzian space-time $(R^4,g)$ is a generic 
counterexample to the {\bf SCCC}} as formulated in the Introduction 
(recall that being generic is not a well-defined concept). We 
also have the suspicion that this particular case sheds light onto the 
general situation i.e., when the space-time is modelled on a 
general non-compact smooth $4$-manifold \cite{gom3, gom-sti}. That is, we 
suspect that the {\bf SCCC} typically fails in four dimensions!


\section{Conclusion and outlook}
\label{three}


From the viewpoint of low dimensional differential topology it is not 
surprising that confining ourselves into the initial value approach when 
thinking about the {\bf SCCC} typically brings affirmative while more 
global techniques might yield negative answers: the initial 
value formulation of Einstein's equations likely just explores 
the vicinity of $3$ dimensional smooth spacelike 
submanifolds inside the full $4$ dimensional space-time. It is 
well-known that an embedded smooth submanifold of an ambient space 
always admits a tubular neighbourhood which is an open disk bundle over 
the submanifold i.e., has a locally product smooth structure. However exotic 
$4$ dimensional smooth structures never arise as products of lower 
dimensional ones consequently the {\it four} dimensional exotica i.e., 
the general structure of space-time never can be detected from a {\it three} 
dimensional perspective such as the initial value formulation. 
There is a {\it qualitative leap} between the two dimensions. 

Finally we make a comment here on Malament--Hogarth space-times and 
``gravitational computers'' as formulated for instance in \cite{ete-nem, 
ete}. Following the terminology introduced in \cite[Definition 3.1]{ete} 
if the maximal Cauchy development of an initial data set is extendible in 
the sense of Definition \ref{perturbacio} then this (necessarily 
non-globally hyperbolic) extension is an example of a {\it generalized 
Malament--Hogarth space-time}; a space-time of this kind is essentially 
conformally equivalent to a {\it Malament--Hogarth space-time} (cf. 
\cite[Remark after Definition 3.1]{ete}). However members of this latter 
class can in principle be used for powerful computations beyond the 
theoretical Turing barrier as explained for instance in \cite{ete-nem} and 
the references therein. Since these space-times are never globally 
hyperbolic (this is well-known, cf. e.g. \cite[Lemma 3.1]{ete}) the {\bf 
SCCC}, if holds, forbids the existence of both physically relevant and stable 
Malament--Hogarth space-times. But our results here demonstrate that 
stable and physically relevant at least {\it generalized} Malament--Hogarth 
space-times exist because the {\bf SCCC} can fail in a generic way. Therefore 
we ask ourselves whether or not our results can be sharpened to prove the 
existence of physically relevant stable Malament--Hogarth space-times: if yes 
then the theoretical possibility of building physically relevant as well as 
stable powerful ``gravitational computers'' would open up \cite{ete-nem, ete}.

\end{document}